\documentclass[aps,prb,twocolumn,showpacs,preprintnumbers,amsmath,amssymb,groupedaddress,superscriptaddress]{revtex4}%

\usepackage{graphicx}
\usepackage{epsfig}
\usepackage{dcolumn}
\usepackage{bm}
\usepackage{longtable}
\usepackage{color}

\begin{document}

\preprint{PREPRINT (\today)}

\title{Rearrangement of the antiferromagnetic ordering at high magnetic fields in SmFeAsO and SmFeAsO$_{0.9}$F$_{0.1}$ single crystals}

\author{S.~Weyeneth}
\affiliation{Physik-Institut der Universit\"at Z\"urich, Winterthurerstrasse 190, CH-8057 Z\"urich, Switzerland}

\author{P.~J.~W.~Moll}
\affiliation{Laboratory for Solid State Physics, ETH Zurich, CH-8093 Zurich, Switzerland}

\author{R.~Puzniak}
\affiliation{Institute of Physics, Polish Academy of Sciences, Aleja Lotnik\'ow 32/46, PL-02-668 Warsaw, Poland}

\author{K.~Ninios}
\affiliation{Department of Physics, University of Florida, Gainesville, Florida 32611, USA}

\author{F.~F.~Balakirev}
\affiliation{National High Magnetic Field Laboratory, Los Alamos National Laboratory, Los Alamos, New Mexico 87545, USA}

\author{R.~D.~McDonald}
\affiliation{National High Magnetic Field Laboratory, Los Alamos National Laboratory, Los Alamos, New Mexico 87545, USA}

\author{H.~B.~Chan}
\affiliation{Department of Physics, University of Florida, Gainesville, Florida 32611, USA}

\author{N.~D.~Zhigadlo}
\affiliation{Laboratory for Solid State Physics, ETH Zurich, CH-8093 Zurich, Switzerland}

\author{S.~Katrych}
\affiliation{Laboratory for Solid State Physics, ETH Zurich, CH-8093 Zurich, Switzerland}

\author{Z.~Bukowski}
\affiliation{Laboratory for Solid State Physics, ETH Zurich, CH-8093 Zurich, Switzerland}

\author{J.~Karpinski}
\affiliation{Laboratory for Solid State Physics, ETH Zurich, CH-8093 Zurich, Switzerland}

\author{H.~Keller}
\affiliation{Physik-Institut der Universit\"at Z\"urich, Winterthurerstrasse 190, CH-8057 Z\"urich, Switzerland}

\author{B.~Batlogg}
\affiliation{Laboratory for Solid State Physics, ETH Zurich, CH-8093 Zurich, Switzerland}

\author{L.~Balicas}\email{balicas@magnet.fsu.edu}
\affiliation{National High Magnetic Field Laboratory, Florida State University, Tallahassee, Florida 32310, USA}

\begin{abstract}
The low-temperature antiferromagnetic state of the Sm-ions in both nonsuperconducting SmFeAsO and superconducting SmFeAsO$_{0.9}$F$_{0.1}$ single crystals was studied by magnetic torque, magnetization, and magnetoresistance measurements in magnetic fields up to 60~T and temperatures down to 0.6~K. We uncover in both compounds a distinct rearrangement of the antiferromagnetically ordered Sm-moments near $35-40$~T. This is seen in both, static and pulsed magnetic fields, as a sharp change in the sign of the magnetic torque, which is sensitive to  the magnetic anisotropy and hence to the magnetic moment in the $ab$-plane, ({\it i.e.} the FeAs-layers), and as a jump in the magnetization for magnetic fields perpendicular to the conducting planes. This rearrangement of magnetic ordering in $35-40$~T is essentially temperature independent and points towards a canted or a partially polarized magnetic state in high magnetic fields. However, the observed value for the saturation moment above this rearrangement, suggests that the complete suppression of the antiferromagnetism related to the Sm-moments would require fields in excess of 60~T. Such a large field value is particularly remarkable when compared to the relatively small N\'{e}el temperature $T_{\rm N}\simeq5$~K, suggesting very anisotropic magnetic exchange couplings. At the transition, magnetoresistivity measurements show a crossover from positive to negative field-dependence, indicating that the charge carriers in the FeAs planes are sensitive to the magnetic configuration of the rare-earth elements. This is indicates a finite magnetic/electronic coupling between the SmO and the FeAs layers which are likely to mediate the exchange interactions leading to the long range antiferromagnetic order of the Sm ions.
\end{abstract}


\maketitle
\section{Introduction}
Soon after the discovery of high-temperature superconductivity in the cuprate system La$_{2-x}$Ba$_{x}$CuO$_4$,\cite{Bednorz} a new class of superconducting compounds based on CuO$_2$ layers was found. Similarly, the recent discovery of superconductivity in LaFeAsO$_{1-x}$F$_y$, with a transition temperature $T_{\rm c}\simeq 26$~K,\cite{Kamihara} led to the discovery of a whole new class of iron-based superconductors  $RE$FeAsO$_{1-x}$F$_y$, where $RE$ denotes a rare earth element, with $T_{\rm c}$'s up to $\simeq 55$~K.\cite{ren} Fe-based superconductors share some common properties with the cuprates such as a layered crystallographic structure, the presence of competing orders, low carrier density, a small coherence length, and possibly also an
unconventional pairing mechanism. As in the cuprates, superconductivity sets in upon doping an antiferromagnetic parent compound.\cite{chen} Nevertheless, there are some important differences: the Fe-based superconductors emerge by doping a metallic parent compound, the anisotropy is in general lower when compared to that of the cuprates, and the symmetry of the order parameter is claimed to be $s\pm$-wave with Fermi-surface nesting playing a major role. \cite{mazin} Therefore, the fundamental question arises whether the mechanisms leading to superconductivity in both families of high temperature superconductors share a common origin.

Neutron scattering experiments revealed that in the undoped compounds the magnetic moments of the Fe ions display a collinear antiferromagnetic order, which is claimed to be itinerant in character.\cite{cruz} It was suggested that the antiferromagnetism in undoped compounds is driven by a nesting instability,\cite{mazin,singh} connecting cylindrical Fermi surfaces of hole and electron character through the antiferromagnetic modulation vector $\vec{Q} =(\pi,\pi)$ (in the original and undistorted tetragonal Brillouin zone),\cite{Ross} hence forming a spin-density wave. Interestingly, the doping dependent phase diagram for the different iron-based superconductors exhibit peculiar features with some compounds exhibiting either coexistence \cite{drew} or competition \cite{cruz2} of antiferromagnetism with superconductivity.\\\indent
The replacement of La in LaFeAsO$_{1-x}$F$_y$ by a magnetic rare earth element as \emph{e.g.} Sm has remarkable consequences: It not only increases the value of $T_{\rm c}$ to 55~K,\cite{ren} but also leads to antiferromagnetic ordering of the rare-earth moments at low temperatures,\cite{ding, Maeter, Ryan, Riggs, Qiu, Tarantini, Kimber, Tian} coexisting in the underdoped compounds with the spin-density wave state due to the magnetic correlation of the Fe ions.\cite{Maeter, Kimber, Tian} At first glance one could expect that the incorporation of ions with a large magnetic moment such as Sm (having a free ion moment of $\mu^{\rm Sm}\simeq0.84~\mu_{\rm B}$)\cite{Ashcroft} to be detrimental for any superconducting pairing scenario.\cite{kontani} However, it has been argued that the main effect of rare-earth elements is related to their ionic radii, and that ions with different radii would change the Fe-As-Fe bond angles and the FeAs-FeAs inter-planar distance.\cite{Vildosola}\\\indent
The occurrence of antiferromagnetic ordering of the Sm magnetic moments has been observed in specific heat experiments in polycrystalline SmFeAsO$_{1-x}$F$_y$.\cite{Ding} The N\'eel temperature was estimated to be $T_{\rm N}\simeq4.6$~K for nonsuperconducting SmFeAsO and $T_{\rm N}\simeq3.7$~K for superconducting SmFeAsO$_{0.85}$F$_{0.15}$. A similar result was found in specific heat investigations at high magnetic fields, where $T_{\rm N}\simeq5.4$~K for SmFeAsO and $T_{\rm N}\simeq3.75$~K for SmFeAsO$_{0.85}$F$_{0.15}$ was deduced.\cite{Riggs} Additionally, the N\'eel temperature of SmFeAsO as derived by the specific heat appears to be almost unaffected by high fields up to 35~T.\cite{Riggs} Neutron diffraction experiments on SmFeAsO reveal at low temperatures a collinear antiferromagnetic ordering of the Sm moments, in which the moments couple ferromagnetically in-plane but antiferromagnetically between adjacent planes with a magnetic moment of $\simeq0.60(3)~\mu_{\rm B}$ per Sm ion.\cite{Ryan} In this regard, SmFeAsO appears to exhibit a type of rare-earth antiferromagnetism which is very similar to the one observed in the layered cuprate system Sm$_2$CuO$_4$.\cite{Sachidanandam, Sumarlin, Strach} This compound was found to display a collinear magnetic-order composed of antiferromagnetically coupled Sm magnetic moments below $T_{\rm N}\simeq6$~K with a magnetic moment of $\simeq0.37(3)~\mu_{\rm B}$ per Sm ion.\cite{Sumarlin} A muon-spin rotation ($\mu$SR) study in SmFeAsO reported a magnetic coupling between the Fe and the \emph{RE} sublattices with an estimated moment of $\simeq 0.4$~$\mu_{\rm B}$ per Sm ion.\cite{Maeter} Intriguingly, in the same work the magnetic structure of the Sm moments was found not only to be non-collinear to the Fe moments, but also to be non-collinear among themselves,\cite{Maeter} refining the picture revealed by the neutron scattering experiment.\cite{Ryan} Interestingly, high-field electron spin resonance spectroscopy in GdFeAsO$_{1-x}$F$_{x}$ also suggest an appreciable exchange coupling between the Gd and Fe moments.\cite{Alfonsov}\\\indent
In order to study the antiferromagnetic state in SmFeAsO$_{1-x}$F$_{y}$ we performed magnetic torque, force magnetometry, and magnetoresistance measurements in single crystals in static magnetic fields up to 45~T and in pulsed fields up to 60~T at $T>0.6$~K. The magnetic torque which initially increases with increasing field exhibits a maximum followed by a sharp reduction in fields of $35-40$~T at low temperatures, and switches its sign at even higher fields. The field at which the torque is maximal displays an angular dependence, different to the one expected for an antiferromagnet in the framework of a classical spin-flop scenario. The distinct signatures of this magnetic behavior, as observed through the different experimental techniques, allow us to gain a deeper insight into the nature of the antiferromagnetic order. Furthermore, as indicated in the experiments, the low temperature antiferromagnetic state of the Sm ions persists to surprisingly high magnetic fields, well beyond the values attained for this study.
\\\indent
\section{Experimental details}
Single crystals of undoped SmFeAsO (crystals A$_1$, A$_2$, A$_3$) and superconducting underdoped SmFeAsO$_{0.9}$F$_{0.1}$ (crystals B$_1$, B$_2$) (all nominal compositions) with masses of a few micrograms were synthesized by the high-pressure cubic anvil technique,\cite{Zhigadlo, Karpinski} and were characterized by X-ray diffraction and by SQUID magnetometry. For the undoped SmFeAsO crystals magnetization studies reveal no traces of superconductivity, whereas for the SmFeAsO$_{0.9}$F$_{0.1}$ single-crystals bulk superconductivity was observed with an average $T_{\rm c}\simeq17$~K, by four-probe resistance measurements performed on single crystals from the same batch. Following the phase diagram published in Ref. \onlinecite{drew} a $T_{\rm c}\simeq17$~K places this compound within the underdoped regime. The few reports on the antiferromagnetic phase-diagram of the Sm ions, indicate a rather weak effect of the F doping on either the N\`{e}el temperature or in the heat capacity anomaly at the transition \cite{ding}.
\\\indent
Torque magnetometry of SmFeAsO and SmFeAsO$_{0.9}$F$_{0.1}$ single crystals at high magnetic fields was performed using static and pulsed-field facilities at the National High Magnetic Field Laboratories (NHMFL) in Tallahassee and Los Alamos, respectively. The crystals were mounted onto piezoresistive silicon micro-cantilevers (SEIKO Instruments, PRC-120 and PRC-400) and measured in a Wheatstone bridge configuration. The sensors with the mounted crystals were placed into $^3$He cryostats capable of achieving temperatures as low as $0.5-0.6$~K. Steady magnetic fields up to 35~T were generated by a Bitter-type resistive coil and up to 45~T by a Hybrid coil-setup magnet. Pulsed magnetic fields up to 60~T were generated by a composite-material solenoid immersed in liquid nitrogen.\\\indent
Low temperature magnetization measurements of antiferromagnetic SmFeAsO along the $c$-axis were performed with a Si based micro-electromechanical capacitive device similar to the one used in Refs.~\onlinecite{Aksyuk} and \onlinecite{bolle}. At low temperatures, the sensor was calibrated by measuring the electrostatic force between the capacitive plates as a function of an external DC-bias voltage.\\\indent
Four-probe magnetoresistance measurements on underdoped SmFeAsO$_{0.9}$F$_{0.1}$ were performed for the current flowing perpendicular to the FeAs planes in pulsed magnetic fields up to 60~T. The sample was prepared using focused ion beam techniques (FIB), with a fabrication process leading to a sample geometry which is identical to the one described in Ref.~\onlinecite{moll}.
\\\indent
\section{Magnetic torque of antiferromagnetic systems}
Antiferromagnetic compounds may show a rich variety of physics at high magnetic fields. While, at low fields the individual magnetic moments prefer to order antiferromagnetically, high magnetic fields may overcome the exchange interaction and reorient the individual magnetic moments, leading in numerous cases to a complex phase diagram with various magnetic field-induced phases. The precise knowledge of the behavior of both the magnetic torque and the magnetization of an antiferromagnetic sample, allows to investigate multiple aspects of magnetic order. Whereas magnetization gives direct information on the magnetic moment oriented along the field, the magnetic torque directly probes the anisotropy of the susceptibility in magnetically ordered or paramagnetic states.
Throughout this manuscript we use the term ``anisotropy" when referring to the intrinsic magnetic anisotropy of antiferromagnetically ordered states, given by ``hard" and ``easy" magnetization/susceptibility axis, or equivalently to an anisotropic susceptibility tensor.
\\\indent
The magnetic torque $\vec{\tau}$ of a single crystal normalized per unit of volume is defined by the vector product of the magnetization $\vec{M}$  and the magnetic field $\vec{H}$
\begin{equation}\label{eq1}
\vec{\tau}=\mu_0(\vec{M}\times\vec{H}).
\end{equation}
Accordingly, the absolute value $\tau=|\vec{\tau}|$ is given by
\begin{equation}\label{eq2}
\tau=\mu_0MH\sin{\varphi},
\end{equation}
with $\varphi$ the angle between $\vec{M}$ and $\vec{H}$.\\\indent
In the well-known case of an antiferromagnet below its N\'eel temperature $T_{\rm N}$,\cite{Yoshida, Tanaka, WangWang, Tokumoto, Kawamoto} a magnetic torque $\tau$ is expected due to the magnetic anisotropy of the ordered magnetic moments. In the classical case of uniaxial magnetic anisotropy with magnetic anisotropy energy $K_{\rm u}$, the free energy $F$ of the system ensemble can be expressed as the sum of the magnetic and the anisotropy energies \cite{Yoshida, Uozaki}
\begin{equation}\label{eq3}
F=-\frac{1}{2}\left(\chi_\perp\sin^2\phi+\chi_\parallel\cos^2\phi\right)\mu_0H^2-K_{\rm u}\sin^2(\phi-\theta).
\end{equation}
Here, $\theta$ describes the angle between magnetic field and the easy axis, $\phi$ is the angle between the magnetic field and the spin axis, and $\chi_\parallel$ and $\chi_\perp$ are the parallel and perpendicular susceptibilities with respect to the easy axis for $K_{\rm u}>0$. With these assumptions the magnetic torque at constant temperature is given by
\begin{equation}\label{eq4}
\tau=-\frac{\partial F}{\partial \theta}=\frac{1}{2}(\chi_\perp-\chi_\parallel)\mu_0H^2\frac{\sin2\theta}{\sqrt{\lambda^2-2\lambda\cos2\theta +1}},
\end{equation}
with
\begin{equation}\label{eq5}
\lambda=\left(\frac{H}{H_{\rm sf}}\right)^2,
\end{equation}
where $H_{\rm sf}$ is the spin-flop field:
\begin{equation}\label{eq6}
H_{\rm sf}=\sqrt{\frac{2K_{\rm u}}{\chi_\perp-\chi_\parallel}}.
\end{equation}
Accordingly, the torque is expected to increase with increasing field up to $H_{\rm sf}$ where a spin reorientation occurs. In excess of this field either a gradual (flop) or discontinuous (flip) alignment of the magnetic moments along the field is expected.\cite{Blundell} In the case of a spin-flop Eq.~(\ref{eq4}) implies that for small angles $\theta\ll45$~deg also the quantity $\tau/H^2$ is expected to increase up to $H_{\rm sf}$. In the low magnetic field limit $H\ll H_{\rm sf}$, Eq.~(\ref{eq4}) reduces to
\begin{equation}\label{eq7}
\tau=\frac{1}{2}(\chi_\perp-\chi_\parallel)\mu_0H^2\sin2\theta,
\end{equation}
and $\tau/H^2$ results to be field independent. Additionally, one derives the field of maximum torque to depend on the angle $\theta$ as
\begin{equation}\label{eq8}
H_{\rm max}(\theta)=\frac{H_{\rm sf}}{{\sqrt{\cos(2\theta)}}}.
\end{equation}
In the case of a spin-flip, Eq.~(\ref{eq7}) is also expected to hold up to the transition at $H_{\rm sf}$, where the discontinuous spin-flip occurs and the torque is reduced in excess to this field.\\\indent
It would be tempting to expect a spin-flop or a spin-flip transition of the Sm-associated magnetic moments given the arrangement of moments suggested by neutron scattering experiments,\cite{Ryan} and the conclusions reached by specific heat studies of SmFeAsO$_{1-x}$F$_y$ (see Ref.~\onlinecite{Riggs}). Such a field induced spin reorientation was reported in the similar compound EuFe$_2$As$_2$.\cite{Jiang, XiaoSu} However, the antiferromagnetic arrangement in SmFeAsO$_{1-x}$F$_y$ appears to be more complicated. Although neutron diffraction results imply a simple collinear antiferromagnetic ordering of the Sm moments,\cite{Ryan} $\mu$SR investigations reveal evidence for an additional antiferromagnetic coupling between the Sm and the Fe magnetic moments, leading to at least three distinct configurations of the Sm moments.\cite{Maeter} For the latter scenario it is not obvious that the above picture of a classical antiferromagnet is sufficient to fully describe the data.
\section{Experimental results}
\begin{figure}[t!]
\centering
\includegraphics[width = 7.4 cm]{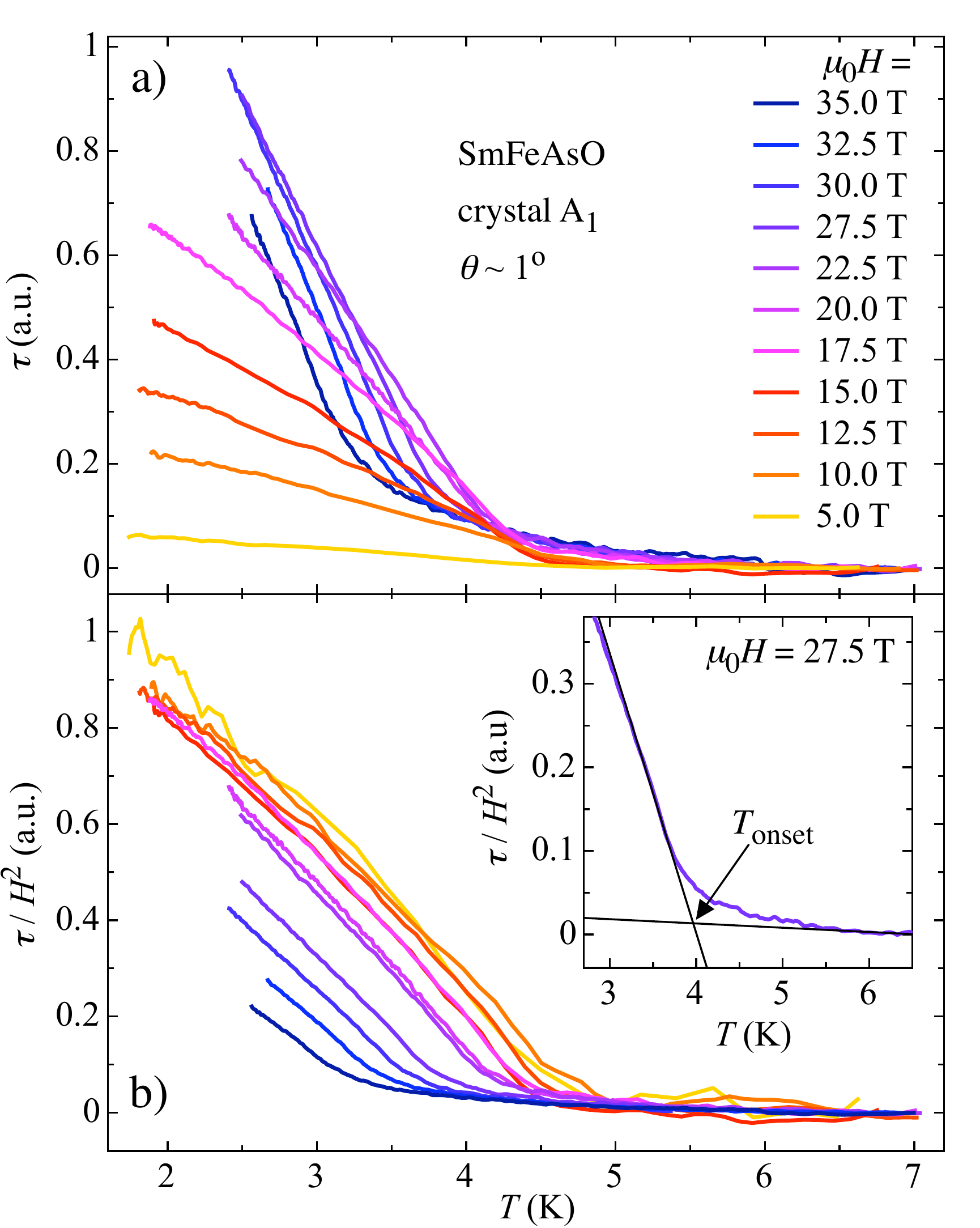}
\caption{(color online) Magnetic torque $\tau$ for a SmFeAsO single crystal in the vicinity of the onset of the antiferromagnetic state for fields nearly along the $c$-axis of the crystal. a) Measured $\tau(T)$ at different magnetic fields $H$ for the SmFeAsO crystal A$_1$, where the angle between $H$ and the $c$-axis is fixed at $\theta\sim1^{\circ}$. b) $\tau/H^2$ as a function of $T$. All branches follow a linear behavior below $T_{\rm onset}(H)$, typical for an antiferromagnetic state. The inset shows $\tau(T)/H^2$ at $\mu_0H=27.5$~T. The temperature $T_{\rm onset}$ is defined by the crossing of the linearly extrapolated data from the high and low temperature regimes, respectively.}
\label{fig1}
\end{figure}
The temperature dependence of the magnetic torque $\tau$ for a SmFeAsO single crystal was measured in various magnetic fields up to 35~T with a fixed angle $\theta\sim1^{\circ}$ and is presented in Fig.~\ref{fig1}a. Below $\sim5$~K a drastic increase in $\tau(T)$ is observed, due to the occurrence of antiferromagnetic order. This is more clearly exposed in Fig.~\ref{fig1}b, where the quantity $\tau/H^2$ is plotted as a function of $T$. All curves show a linear dependence in temperature below a characteristic temperature $T_{\rm onset}(H)$. Notice that $T_{\rm onset}(H)$ decreases with increasing magnetic field. The nearly linear dependence of $\tau/H^2$ (see Fig.~\ref{fig1}b) at low temperatures is typical for an antiferromagnetic state according to Eq.~(\ref{eq4}), assuming $\chi_\perp$ to be temperature independent and $\chi_\parallel$ to scale linearly with $T<T_{\rm N}$ (see \emph{e.g.} Ref.~\onlinecite{Blundell}). The inset to Fig.~\ref{fig1}b shows $\tau/H^2$ measured at $\mu_0H=27.5$~T, demonstrating how $T_{\rm onset}(H)$ is extracted from the data (by analyzing the crossing of two straight lines obtained by extrapolating the low and the high temperature behavior of $\tau/H^2$).\\\indent
\begin{figure}[t!]
\centering
\includegraphics[width = 7.4 cm]{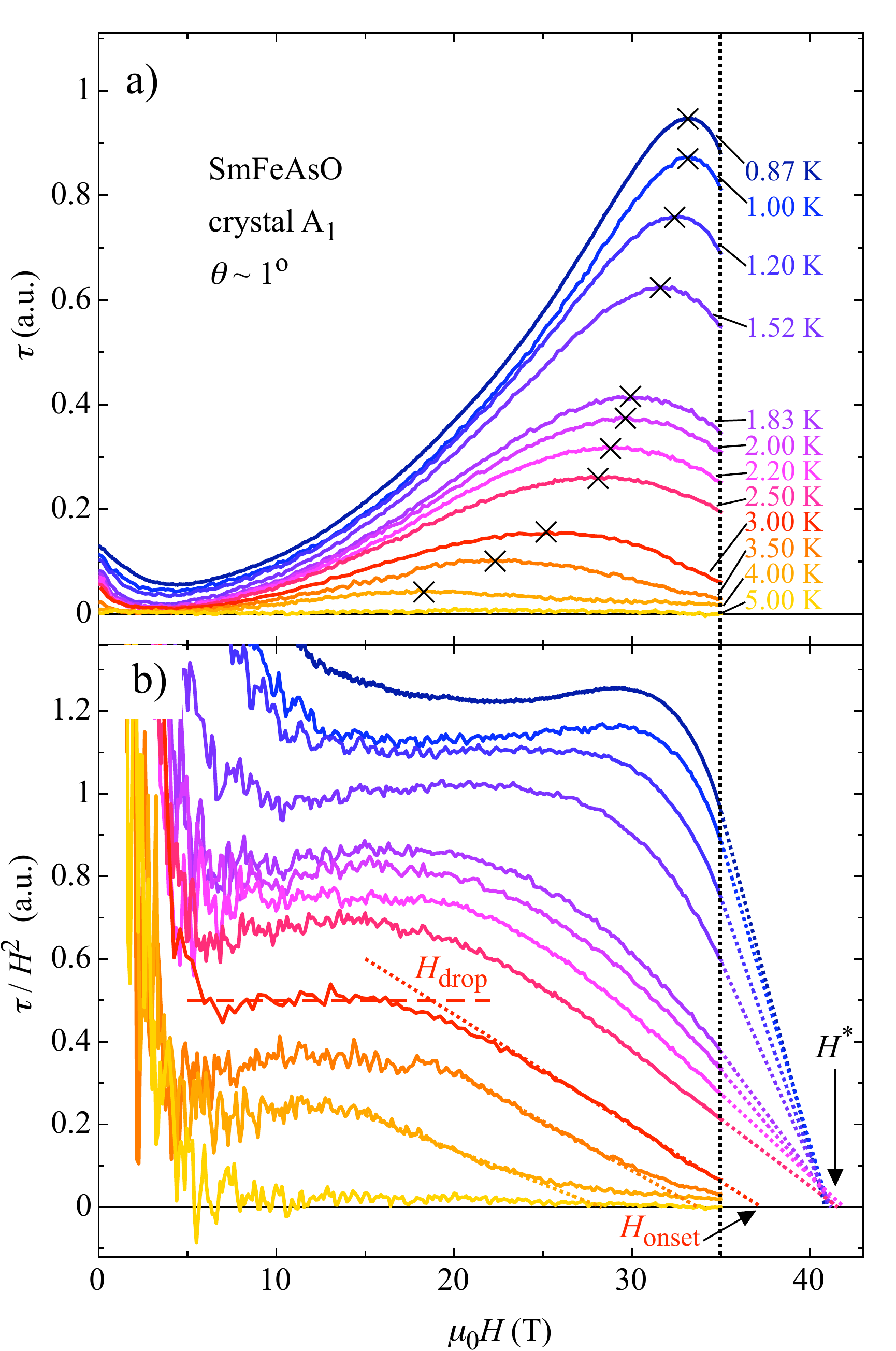}
\caption{(color online) High-field behavior of the magnetic torque in antiferromagnetic SmFeAsO. a) Magnetic torque $\tau(H)$ at different temperatures for the SmFeAsO crystal A$_1$, where the angle between $H$ and the $c$-axis is fixed to $\theta\sim1^{\circ}$. The maximum in torque at the field $H_{\rm max}$ is marked by the crosses and is temperature dependent. b) $\tau/H^2$ as a function of $H$. All curves are almost field independent in the intermediate field range $10-20$~T. For temperatures $T\geqslant3$~K the dotted lines above fields exceeding 35~T are extrapolations to estimate $H_{\rm onset}(T)$, where $\tau(H)/H^2$ is expected to reach zero. Below $T\leqslant2.5$~K, all these extrapolations point to the same field $\mu_0H^*\approx40$~T. $H_{\rm drop}(T)$ denotes the field where $\tau/H^2$ starts being suppressed. The convention for $H_{\rm drop}$ and $H_{\rm onset}$ is shown for the $T=3$~K data.}
\label{fig2}
\end{figure}
\begin{figure}[t!]
\centering
\includegraphics[width = 7.4 cm]{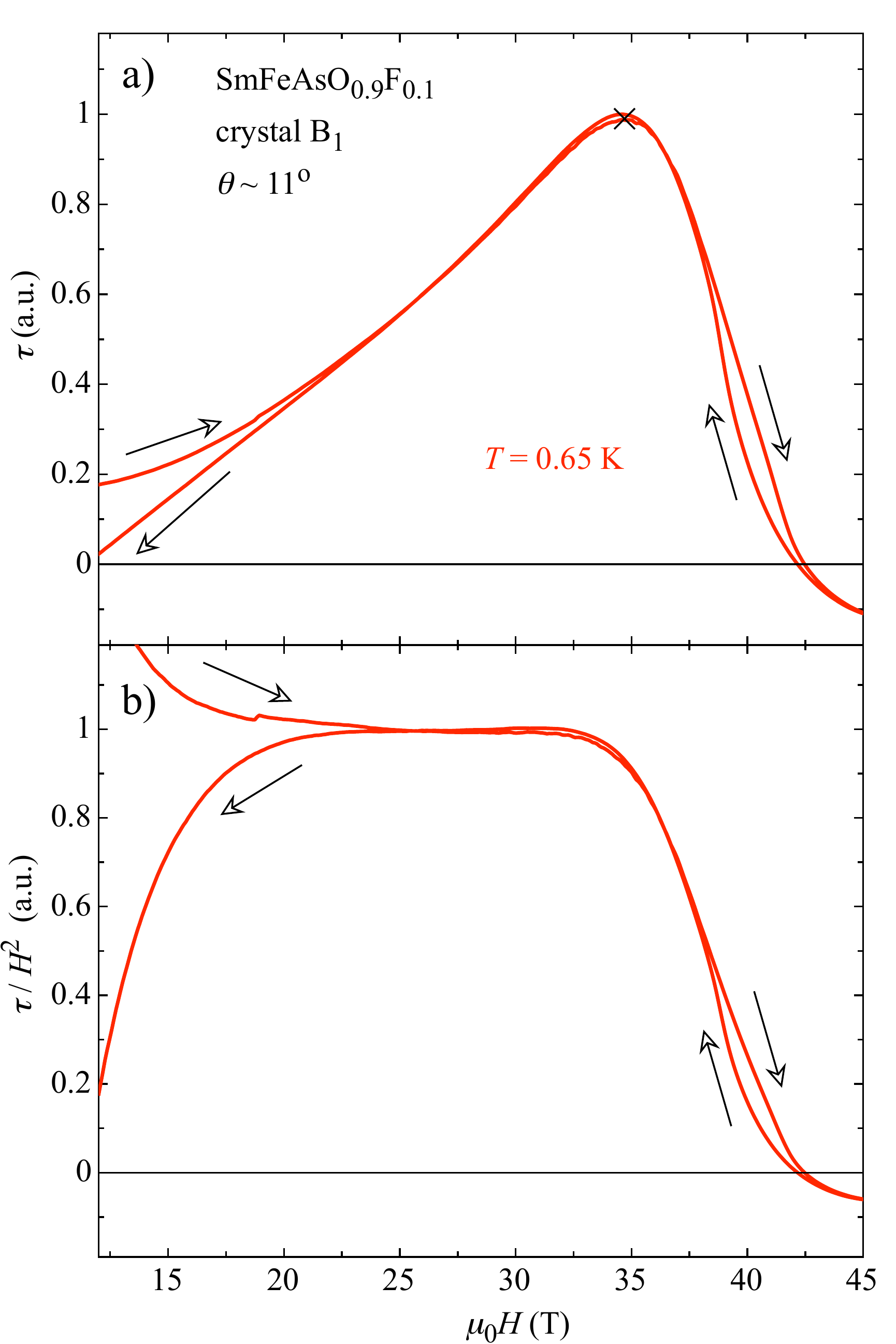}
\caption {(color online) Magnetic torque $\tau$ as a function of $H$ for a superconducting SmFeAsO$_{0.9}$F$_{0.1}$ single crystal. a) $\tau$ for crystal B$_1$ ($T_{\rm c}\simeq17$~K) at a fixed angle $\theta\simeq11^{\circ}$ and at 0.65 K. The torque maximum is marked by a cross. A hysteresis is observed above the maximum in $\tau(H)$. b) $\tau/H^2$ as a function of $H$.}
\label{fig3}
\end{figure}
Figure~\ref{fig2}a shows the magnetic field dependence of the torque signal $\tau(H)$ for a SmFeAsO single crystal in the temperature range $T\leqslant5$~K, measured at a fixed angle $\theta\sim1^{\circ}$. Whereas at $T\sim5$~K essentially no torque signal is observed, $\tau(H)$ increases strongly with decreasing temperature. At very low magnetic fields, the magnetic torque in the antiferromagnetic state is expected to be almost zero, since here $\tau(H)$ is proportional to $H^2$. However, the data presented in Fig.~\ref{fig2}a, indicate an additional contribution to the torque at low fields, possibly due to a change in the antiferromagnetic anisotropy at low magnetic fields, or some small additional anisotropic magnetic contributions to the torque. Above $\mu_0H\sim5$~T, $\tau(H)$ increases almost quadratically with $H$ towards a field dependent maximum beyond which it exhibits a pronounced decrease, indicating a change in the original spin arrangement at very high fields. We define the magnetic field where the maximum in $\tau(H)$ is observed as the field $H_{\rm max}(T)$ which is indicated by crosses in Fig.~\ref{fig2}a. In Fig.~\ref{fig2}b the quantity $\tau/H^2$ is plotted as a function of $H$. At intermediate field strengths, $\tau/H^2$ is essentially constant as a function of $H$, whereas at low magnetic fields, the previously discussed additional contribution dominates $\tau/H^2$ which tends to diverge as the field is ramped down to zero. Above $H_{\rm drop}$ the quantity $\tau/H^2$decreases rapidly with increasing $H$. The linearly extrapolated data, as shown by the dotted lines in Fig.~\ref{fig2}b, leads to an estimate of the magnetic field where $\tau$ becomes zero. Down to $3$~K these fields are found to be strongly temperature dependent and are denoted as $H_{\rm onset}(T)$, whereas the estimated fields for $T\leqslant2.5$~K are almost constant and are labeled as $H^*(T)$. Clearly $H_{\rm onset}$ scales with the value of $T_{\rm onset}$ as defined in Fig.~\ref{fig1}b.\\\indent
\begin{figure}[t!]
\centering
\includegraphics[width = 7.4 cm]{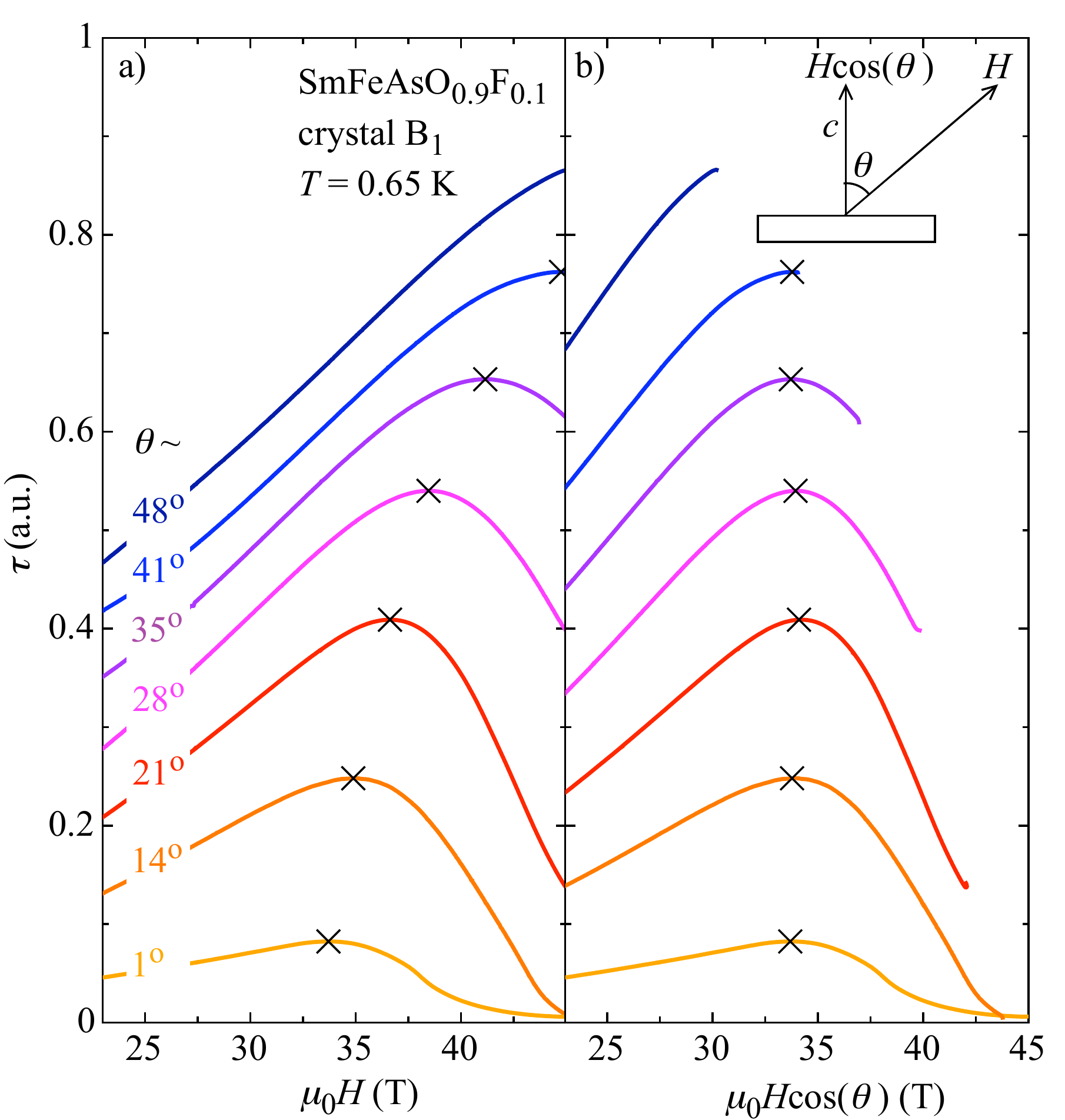}
\caption{(color online) Magnetic torque $\tau(H)$ as a function of the magnetic field $H$ and for various angles $\theta$ between $H$ and $c$-axis of the superconducting SmFeAsO$_{0.9}$F$_{0.1}$ single crystal B$_1$ ($T_{\rm c}\simeq17$~K). The maxima shift with $\theta$ according to $H/\cos(\theta)$ (panel a) or, equivalently, occur at the same $c$-component $H\cos(\theta)\simeq34$~T perpendicular to the planes (panel b). Hence, the position of the torque maximum observed at high fields is related to a magnetization component in the $ab$-plane.}
\label{fig4}
\end{figure}
The data presented in Figs.~\ref{fig1} and \ref{fig2} suggest the occurrence of a metamagnetic transition in SmFeAsO which might correspond to either an onset of a gradual spin-reorientation or a discontinuous spin-canting transition at the field $H_{\rm max}$. Throughout this manuscript we define the term ``metamagnetism" simply as a superlinear increase in the magnetization under a given external field \cite{perry}. As this anomaly in magnetic torque is detected in the range of temperatures where the antiferromagnetism of Sm-ions is present, one might expect a similar transition for superconducting SmFeAsO$_{0.9}$F$_{0.1}$ single crystals. Notice that the Sm-associated antiferromagnetic order was observed in superconducting SmFeAsO$_{1-x}$F$_{y}$ samples.\cite{Riggs} In Fig.~\ref{fig3}a we show $\tau(H)$ for the superconducting SmFeAsO$_{0.9}$F$_{0.1}$ single crystal B$_1$ at 0.65~K and at an angle $\theta\sim11^{\circ}$, acquired by increasing and decreasing the external magnetic field from $\mu_0H=11.5$~T to 45~T, respectively. Both, increasing and decreasing field branches of $\tau(H)$ are displayed in order to visualize the irreversible response characteristic of the superconducting state below the irreversibility field $H_{\rm irr}$ and hysteretic behavior observed for fields beyond $H_{\rm max}$. For this sample with a rather low $T_{\rm c}\simeq17$~K, we do not expect $H_{\rm irr}$ to be located above $\simeq23$~T, even at 0.6~K, as indicated by the low field irreversible branches in Fig.~\ref{fig3}a. Hence, the irreversible torque signal above $H_{\rm max}$ is not related to superconductivity, but to the antiferromagnetic state of the Sm magnetic moments. The data suggests that the torque would display negative values for fields exceeding a certain value $H^*$. In Fig.~\ref{fig3}b we present the same data set but as $\tau(H)/H^2$ as a function of $H$ which displays very similar qualitative behavior with respect to the traces presented in Fig.~\ref{fig2}b. By comparing Figs.~\ref{fig2}a and \ref{fig3}a, one notices that the maximum in the torque signal at low temperatures occurs at a slightly higher magnetic field $H_{\rm max}\simeq35$~T for the superconducting crystal B$_1$ than the one for the nonsuperconducting SmFeAsO crystal A$_1$ with $H_{\rm max}\simeq33$~T. However, the data of each sample were recorded at different angles $\theta$. Hence, in order to compare the data presented in Figs.~\ref{fig2} and \ref{fig3}, we investigated the angular dependence of $H_{\rm max}$ in the SmFeAsO$_{0.9}$F$_{0.1}$ crystal B$_1$. Figure~\ref{fig4}a shows the magnetic torque $\tau(H)$ for various angles $\theta$. The maxima in $\tau(H)$ are clearly shifting to higher fields with increasing $\theta$. Figure~\ref{fig4}b shows $\tau$ as a function of the rescaled magnetic field
\begin{equation}\label{eq9}
H^{||c}=H\cos(\theta).
\end{equation}
All torque maxima in $\tau(H^{||c})$ are observed at essentially the same $c$-axis component of the magnetic field. This would suggest that the maximum of torque is associated with a reorientation or canting of a magnetization component in the conducting planes.\\\indent
\begin{figure}[t!]
\centering
\includegraphics[width = 7.4 cm]{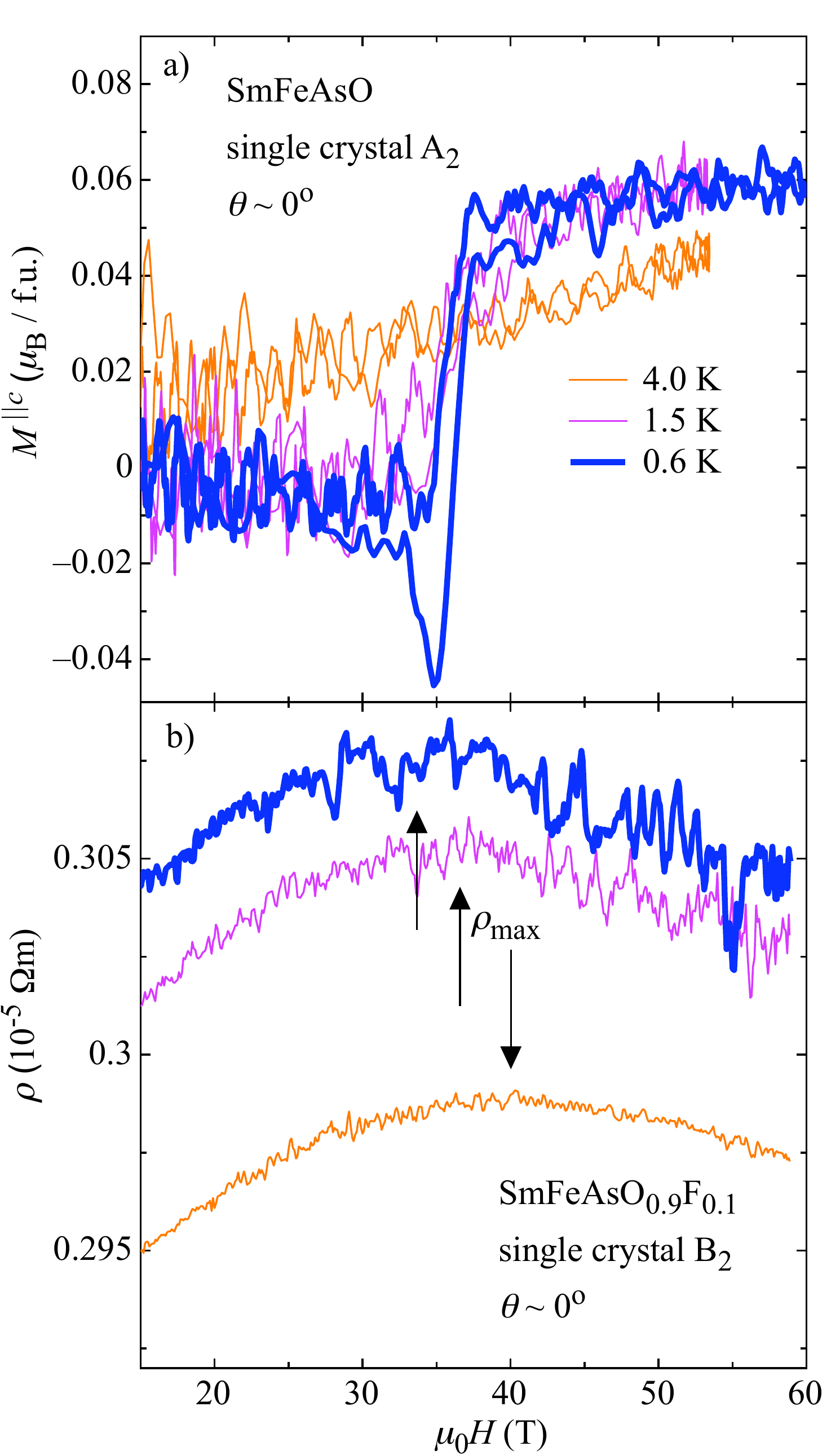}
\caption {(color online) Experimental indication for a rearrangement of the $c$-axis component of the magnetization at high fields. a) Magnetization measured with the field perpendicular to the planes in a force magnetometer for the SmFeAsO crystal A$_2$ at $T \simeq0.6$~K, $\simeq1.5$~K, and $\simeq4$~K, respectively. At the tempeture $T \simeq0.6 K$ the magnetization jumps by $M_{\rm jump}\simeq0.06(2)$~$\mu_{\rm B}$ per formula unit (f.u.) at $\sim35$~T. b) Magnetoresistance $\rho(H)$ of the SmFeAsO$_{0.9}$F$_{0.1}$ crystal B$_2$ in pulsed magnetic fields perpendicular to the planes. The magnetoresistance shows a maximum $\rho_{\rm max}$ between $35-40$~T, which shifts to higher fields with increasing temperatures (indicated by the arrows).}
\label{fig5}
\end{figure}
Obviously, the observed angular dependence of the torque maxima is different from that expected for a classical anisotropic antiferromagnet [see Eq~(\ref{eq8})]. However, both SmFeAsO and SmFeAsO$_{1-x}$F$_y$ exhibit a similar response in the torque concerning the rearrangement of the low-temperature antiferromagnetic order in fields up to $\sim35$~T, and show a clear drop in $\tau(H)$ above $H_{\rm max}(T)$. Clearly, only the $c$-axis component of the magnetic field is responsible for this distinct feature in the torque signal.
\\\indent
We extracted the absolute value of the change in the $c$-axis component of the magnetic moment $m_z$ at $H_{\rm max}$ by performing magnetic force measurements at low temperatures. The mass of the crystal A$_2$ which was selected from the same batch of SmFeAsO investigated above, was determined by monitoring the shift in the resonance frequency of the device and was found to be equal to $\simeq2.1(6)~\mu$g. The sensor with the mounted sample was placed slightly off field center, in order to make use of the magnetic field gradient $\partial H/\partial z$ of the solenoid. The value of $\partial H_z/\partial z$ was estimated by analyzing the geometry of the solenoid. The magnetization along the $c$-axis is then derived accordingly
\begin{equation}
M_z =\frac{F_z}{\mu_0V}\left(\frac{\partial H_z}{\partial z}\right)^{-1}.
\end{equation}
The change in magnetization at $H_{\rm max}$ is found by extracting the magnetic moment from magnetic force measurements performed as a function of the field applied along the $c$-axis. Figure \ref{fig5}a presents the calibrated magnetization data obtained at various temperatures. As clearly seen, the magnetization shows a sharp jump at $\simeq35$~T at $T=0.6$~K ($\theta$ is kept at nearly zero degrees during these measurements) saturating at a value of only $\sim0.06(2)~\mu_{\rm B}$ per formula unit (f.u.). This value is rather small, when compared to the estimates for the full Sm magnetic moment of $\simeq0.4-0.6~\mu_{\rm B}$ reported in the literature,\cite{Maeter, Ryan} suggesting that only a partial reorientation of the magnetic moments is observed at $H_{\rm max}\simeq35$~T. These results indicate that much higher fields are required to fully suppress the antiferromagnetic order. Notice that no additional jumps  in the $c$-axis component of the magnetization are observed at base temperature and under fields up to 60~T, see Fig.~\ref{fig5}a.\\\indent
\begin{figure*}[t!]
\centering
\includegraphics[width = 15 cm]{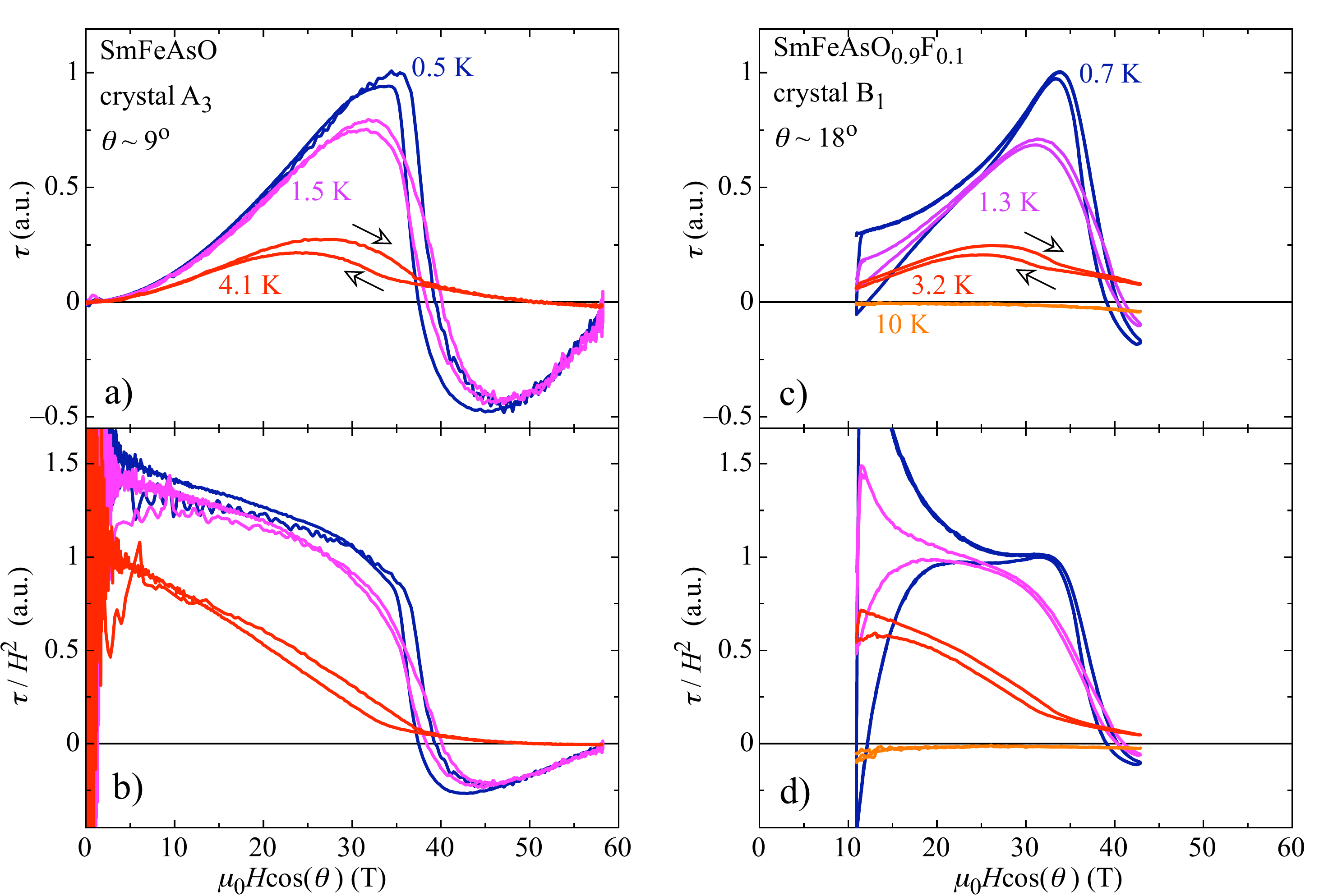}
\caption{(color online) Magnetic torque for single crystalline SmFeAsO and SmFeAsO$_{0.9}$F$_{0.1}$. For a comparison with measurements perfomed at different angles $\theta$ (see Figs.~\ref{fig2}, \ref{fig3}, and \ref{fig4}) the field is given as $H\cos(\theta)$. a) $\tau(H)$ for the SmFeAsO crystal A$_3$ measured at a fixed angle $\theta\sim9^{\circ}$. b) $\tau/H^2$ obtained from the data shown in panel a). c) and d) the same as in a) and b) but for the  SmFeAsO$_{0.9}$F$_{0.1}$ single crystal B$_1$.}
\label{fig6}
\end{figure*}
We have also studied the magnetoresistance $\rho(H)$ of the underdoped SmFeAsO$_{0.9}$F$_{0.1}$ crystal B$_2$ for electric currents flowing perpendicularly to the FeAs-planes in pulsed magnetic fields up to 60~T. These experiments provide additional evidence for the observed magnetic rearrangement (see Fig.~\ref{fig5}b). At fields below $35-40$~T, a positive magnetoresistance was observed, while the material shows negative magnetoresistance at higher fields. In our scenario, the negative magnetoresistance is indicative of a reduction of spin scattering as the Sm ions undergo a spin reorientation transition. The maximum observed in the $\rho(H)$ curves shifts to higher fields with increasing temperature. At low temperatures, the field where the maximum $\rho_{\rm max}$ in the resistivity is observed, coincides with the field where the jump in magnetization occurs.\\\indent
In order to further characterize this high-field metamagnetic behavior associated with the Sm-antiferromagnetic order, we performed additional torque measurements in pulsed magnetic fields up to 60~T for the SmFeAsO crystal A$_3$. Figure~\ref{fig6}a displays $\tau(H)$ recorded at various temperatures and at a fixed angle $\theta\sim9^{\circ}$. At fields below $35-40$~T the qualitative behavior of $\tau(H)$ strongly resembles the data shown in Fig.~\ref{fig2}a. It is found that for $H>H^*$ the magnetic torque indeed becomes negative. It is interesting to note that a change of sign in magnetic torque could also be related to a change of easy axis in magnetic ordering. A similar effect was reported for molecular magnets.\cite{Waldmann} In Fig.~\ref{fig6}b $\tau(H)/H^2$ is presented. The $c$-axis components of the fields $H_{\rm max}$ and $H^*$ as observed in Figs.~\ref{fig6}a and \ref{fig6}b are in terms of $\tau(H\cos(\theta))$ in good agreement to those determined in Figs.~\ref{fig2}a and \ref{fig2}b. Figures \ref{fig6}c and \ref{fig6}d show magnetic torque data obtained for the SmFeAsO$_{0.9}$F$_{0.1}$ crystal B$_1$ probed by static magnetic field. Obviously, the magnetic torque for both, superconducting and non-superconducting samples exhibit the same qualitative behavior in field.\\\indent

Reviewing the overall experimental data for undoped SmFeAsO presented in Figs.~\ref{fig1}, \ref{fig2}, \ref{fig5}, and \ref{fig6} one concludes that the original antiferromagnetic arrangement of the Sm-magnetic moments persists to high magnetic fields of the order of 35~T. Although $T_{\rm N}$ is only about $5$~K, it is remarkable that an energy scale of 35~T is required to perturb this antiferromagnetic ground state. This scenario can be explained by invoking a difference in the energies corresponding to the ordering temperature $T_{\rm N}$ and the spin-flop field $H_{\rm sf}$, resulting from a reduced dimensionality in the magnetic interactions. In this picture the magnetic exchange constants, characterized most likely by an in-plane exchange constant $J$ and a much smaller out of the plane one $J_{\perp}$, could be very anisotropic, similar to the situation in Ref.~\onlinecite{Goddard}. A phase-transition takes place, when all moments become coherently coupled below $k_{\rm B}T_{\rm N}\propto J_{\perp}\ll J$. However, the metamagnetic transition, or the field-induced rearrangement of an anisotropic antiferromagnetic configuration of localized moments as observed here, would involve all the relevant energy scales leading to such a configuration, in particular the largest $J$, thus explaining the large value of the saturation field $H_{\rm sf}\propto J\gg J_{\perp}$. As for undoped SmFeAsO more than two anisotropic exchange energies might be involved in the antiferromagnetic order, as we may infer from results of earlier investigations.\cite{Riggs, Maeter}\\\indent
\begin{figure}[t!]
\centering
\includegraphics[width = 7.4 cm]{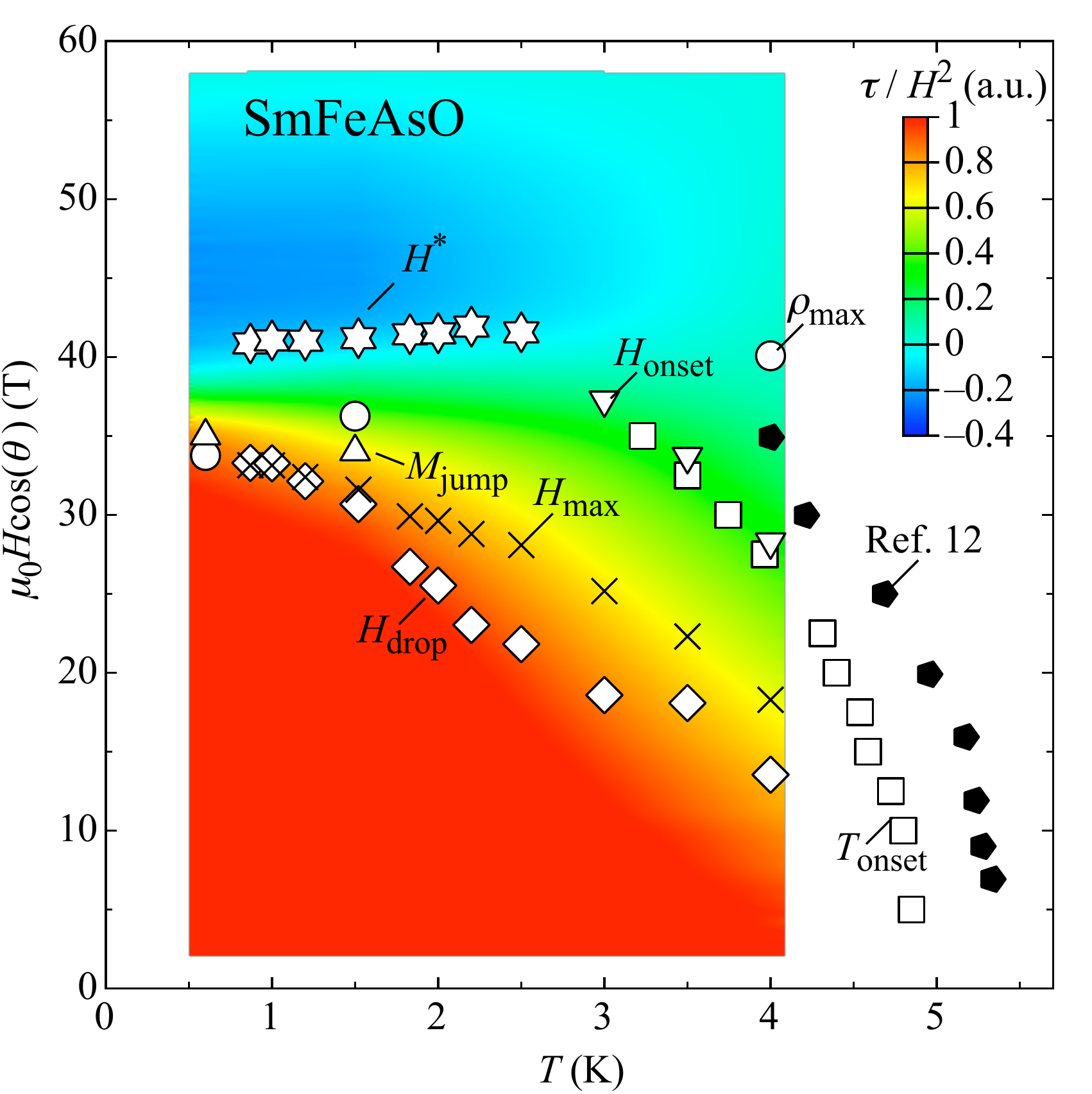}
\caption{(color online) $H-T$ phase-diagram of SmFeAsO from magnetic torque $\tau$, magnetization $M$, and magnetoresistivity $\rho$ measurements. At low temperatures, the application of high magnetic fields in the order of 35 to 40~T, disturbs the original antiferromagnetic state of the Sm-ions resulting in an antiferromagnetic rearrangement in fields exceeding $H_{\rm max}$. For the sake of comparison, we include the phase-diagram obtained from specific  heat measurements in polycrystalline samples (having a higher N\`{e}el temperature $T_{\rm N}\simeq5.4$~K).\cite{Riggs} The different symbols depict the field and temperature dependence of various physical quantities defined in previous figures and within the text.}
\label{fig7}
\end{figure}
In Fig.~\ref{fig7} a color map of $\tau(H)/H^2$ as a function of $\mu_0H\cos(\theta)$ and $T$, based on the results for SmFeAsO, is presented. The experimental data shown in Fig.~\ref{fig6}b have been combined in order to generate a map, illustrating $\tau(H)/H^2$ in a normalized scale as a function of temperature and field. For completeness our estimates of the quantities $H_{\rm max}$, $H_{\rm drop}$, $H^*$, $H_{\rm onset}$, $T_{\rm onset}$, $M_{\rm jump}$, and $\rho_{\rm max}$ of Figs.~\ref{fig1}, \ref{fig2}, and \ref{fig5} are shown as well. Interestingly, the region where the change of the sign of the torque is observed, appears to be almost temperature independent.\\\indent
\section{Discussion and Summary}
In summary in this work, we present evidence for a metamagnetic anomaly occurring in magnetic fields in the order of $35-40$~T in both SmFeAsO and SmFeAsO$_{0.9}$F$_{0.1}$ single crystals. At first glance, it would seem that this transition corresponds to the suppression of the antiferromagnetic order of the Sm-ions. However, we found the change of the saturation moment at the transition [$\simeq0.06(2)~\mu_{\rm B}$] to be much smaller than the value for the full Sm-moments of $\simeq0.4-0.6~\mu_{\rm B}$ extracted by $\mu$SR and neutron diffraction.\cite{Maeter, Ryan} Hence, although the observed anomaly in the torque in magnetic fields between $H_{\rm max}(T)$ and $H_{\rm drop}(T)$ is obviously related to a spin-canting, evidenced by the similarity of the presented data to the expectations for a spin-flop or a spin-flip, it apparently involves a partial spin reorientation only. This scenario appears to be in agreement to the $\mu$SR result of SmFeAsO, which suggests a rather complex magnetic structure of the sublattice of Sm magnetic moments, where the spins are not expected to align collinearly. Our results indicate that extended high magnetic fields investigations in SmFeAsO are required to reach the full polarization of the magnetic moments at low temperatures. It remains to be answered in which manner antiferromagnetism in SmFeAsO$_{1-x}$F$_y$ is suppressed in even higher fields. It is possible that the spin density wave state formed due to the Fe magnetic moments is crucial for this high magnetic field behavior.\\\indent
We show evidence for the existence of a rearrangement of antiferromagnetic ordering in the SmFeAsO$_{1-x}$F$_y$ series, which is undoubtedly associated with the N\'{e}el state of the Sm-moments and their coupling to the Fe sublattice. Extremely high magnetic fields are necessary to induce it, despite the relatively low value of the N\'{e}el temperature, indicating that this antiferromagnetic state is highly anisotropic, {\it i.e.} quasi-two-dimensional. The very low value of the recovered saturation moment suggests that a new antiferromagnetic state is induced by high magnetic fields with partially reoriented or canted magnetic moments. A full suppression of Sm antiferromagnetism would require enormous magnetic fields which are beyond the field range of this study.\\\indent
It might be worthwhile to further investigate under high magnetic fields the detailed magnetic and superconducting phase diagram of the various $RE$FeAsO$_{1-x}$F$_y$ systems, in order to elucidate in greater detail the role of the interaction between the magnetic moment of the rare-earth ion and the electrons mainly responsible for the superconducting state. An obvious route would be to fully explore the upper critical field $H_{\rm c2} (T)$, which, due to its large value, is only partially accessible to the present study. Apparently, there might be some differences in the temperature evolution of $H_{\rm c2}$ between the SmFeAsO$_{1-x}$F$_y$ (Refs.~\onlinecite{Karpinski} and \onlinecite{moll}) and the NdFeAsO$_{1-x}$F$_y$ (Ref.~\onlinecite{Jaroszynski}) compounds. Only further studies, perhaps not based solely on dissipative transport measurements, can clarify the origin of such differences.\\\indent
Finally, and although the incorporation of magnetic rare-earth elements in the SmFeAsO$_{1-x}$F$_y$ system increases the superconducting transition temperature considerably, our observations suggest that its magnetism may be detrimental for superconductivity at very high magnetic fields. The clarification of this point is particularly relevant for these materials, since the combination of extremely high upper critical fields,\cite{Jaroszynski, moll} large and isotropic critical currents,\cite{moll, Zhigadlo, Karpinski, Zhigadlo2} and unconventional magnetic anisotropy,\cite{Weyeneth1, Weyeneth2} make them potentially relevant for technological applications.\\\indent

\section{Acknowledgements}
The authors acknowledge stimulating discussions to A.~Gurevich and F.~Mila. The NHMFL is supported by NSF through NSF-DMR-0084173 and the State of Florida. L.~B. is supported by DOE-BES through award DE-SC0002613. K.~N., H.~B.~C., and L.~B. are supported by the NHMFL-UCGP program. This work was also partially supported by the Swiss National Science Foundation, the NCCR program MaNEP, and the Polish Ministry of Science and Higher Education, within the research project for the years 2007-2010 (No.~N~N202~4132~33).

\end{document}